\documentstyle[twocolumn,aps,psfig]{revtex}

\begin{document}

\draft
\flushbottom
\twocolumn[
\hsize\textwidth\columnwidth\hsize\csname @twocolumnfalse\endcsname

\title{Relaxation effects in the charge-ordered state of La$_{0.5}$Ca$_{0.5}$MnO$%
_{3}$}
\author{V. N. Smolyaninova, C. R. Galley, and R. L. Greene}
\address{Center for Superconductivity Research,\\
Department of Physics, University of Maryland, College~Park,\\
Maryland 20742-4111 }
\date{\today}
\maketitle
\tightenlines
\widetext
\advance\leftskip by 57pt
\advance\rightskip by 57pt

\begin{abstract}
We report an experimental study of the time dependence of the resistivity
and magnetization of charge-ordered La$_{0.5}$Ca$_{0.5}$MnO$_{3}$ under
different thermal and magnetic field conditions. A relaxation with a
stretched exponential time dependence has been observed at temperatures
below the charge ordering temperature. A model using a hierarchical
distribution of relaxation times can explain the data.

\end{abstract}

]
\narrowtext

\tightenlines

Recently, mixed-valent perovskite manganese oxides (A$_{1-x}$B$_{x}$MnO$_{3}$%
: A = La, Pr, Nd; B = Ca, Sr, Ba) have stimulated a great deal of research
interest due to their remarkable electronic and magnetic properties \cite
{rev1,rev2}. An interplay among the charge carriers, magnetic coupling,
orbital ordering and structural distortion leads to a variety of physical
properties ranging from ferromagnetic (FM) metal to antiferromagnetic (AFM)
charge-ordered (CO) insulator depending on the concentration of doping ($x$%
), temperature, and magnetic field. According to the phase diagram \cite
{rev2}, at low temperatures La$_{1-x}$Ca$_{x}$MnO$_{3}$ is a FM metal for $%
0.17<x<0.5$ and an AFM charge-ordered insulator for $0.5<x<0.88$. The region
of Ca doping in the vicinity of $x=0.5$ is of special interest because a 
{\it commensurate }(1:1) charge ordering of Mn$^{3+}$ and Mn$^{4+}$ ions
occurs at low temperatures. This charge ordering can be destroyed
(``melted'') by the application of a modest magnetic field, resulting in a
FM metallic state \cite{Kuwahara,COmelt,Xiao}. Surprisingly, the strength of
this ``melting'' field is an order of magnitude smaller than the charge
ordering temperature (T$_{CO}$) on the energy scale. Also the CO manganites
(at $x=0.5$) exhibit a memory effect \cite{Xiao,Kiryukhin}. For example,
once driven into a metallic state by application of a magnetic field, La$%
_{0.5}$Ca$_{0.5}$MnO$_{3}$ tends to retain the metallic state even if the
magnetic field is removed \cite{Xiao}. Moreover, a fine balance between the
kinetic energy and the Coulomb repulsion of the charge carriers in the
presence of magnetic field results in an electronic phase separation (the
coexistence of the FM metallic and AFM CO phases) for certain values of
magnetic field and temperatures \cite{Kuwahara}. A detailed understanding of
the ``melting'' of the CO state and the electronic phase separation is
lacking at present. In an effort to better understand the dynamics of the
competing interactions in charge-ordered La$_{0.5}$Ca$_{0.5}$MnO$_{3}$, we
present here a study of the relaxation of the magnetization and resistivity
under various field and temperature conditions.

Polycrystalline samples of La$_{0.5}$Ca$_{0.5}$MnO$_{3}$ and La$_{0.47}$Ca$%
_{0.53}$MnO$_{3}$ were prepared from stoichiometric amounts of La$_{2}$O$%
_{3} $, CaCO$_{3}$ and MnCO$_{3}$ by a standard solid-state reaction
technique. X-ray powder diffraction and neutron diffraction \cite{Lynn} show
a single-phase structure with no detectable impurity phases. Magnetization
was measured with a commercial SQUID magnetometer, and resistivity was
measured by a standard 4-probe technique.

At the charge ordering temperature (T$_{CO}=150$ K) the resistivity of La$%
_{0.5}$Ca$_{0.5}$MnO$_{3}$ increases sharply to a large, insulating value at
low temperature (Fig. 1a, H = 0 curve). From magnetization and neutron
scattering \cite{Lynn} we find that the onset of the charge ordering is
accompanied by a transition to the antiferromagnetic state. As shown in Fig.
1a, application of a 5 T magnetic field to zero field cooled (ZFC) La$_{0.5} 
$Ca$_{0.5}$MnO$_{3}$ has little effect on the resistivity. On the contrary,
if the sample is cooled in a magnetic field of 5 T, the resistivity of the
sample is lower than the zero field value by several orders of magnitude.
This suggests that the charge ordering is partially destroyed by cooling
down in a magnetic field. All these results are in agreement with prior work 
\cite{Xiao}.

The time dependencies of the resistivity and magnetization were studied
after the following types of magnetic and thermal treatments: Path 1 - the
sample was cooled in a magnetic field, then the field was decreased to zero;
Path 2 - the sample was cooled in zero magnetic field, then the magnetic
field was increased to the required value; Path 3 - the sample was cooled in
zero magnetic field, the magnetic field was applied, then the magnetic field
was decreased to zero.

In Fig. 1b we show the time dependence of the resistivity ($\rho (t)$) of La$%
_{0.5}$Ca$_{0.5}$MnO$_{3}$ taken after the sample was cooled to 12 K in a
field of 5 T and the field was then switched off (Path 1). The data were
taken during approximately 27 hours. The change in resistivity during this
process from the starting point (1) to the ending point (2) is shown in Fig.
1a as 1 $\rightarrow $ 2. The field cooled (FC) state appears to be a
metastable state, because the resistivity does not return to the zero field
value when the magnetic field is switched off, but rather relaxes slowly
with time nonexponentially. A logarithmic time dependence can fit some
portions of the relaxation curve, but cannot fit the entire $\rho (t)$
curve. We found that the time dependence of the resistivity can be fit with
a stretched exponential dependence (Fig. 1b): 
\begin{equation}
\rho (t)=\rho _{2}+(\rho _{1}-\rho _{2})exp[-(t/\tau )]^{\beta }
\end{equation}
where $\rho _{1}=446\pm 2$ ohm-cm, $\rho _{2}=2.135\pm 0.005$ ohm-cm, $\tau
=34700\pm 400$ sec, and $\beta =0.330\pm 0.001$.

\begin{figure}[tbp]
\centerline{
\psfig{figure=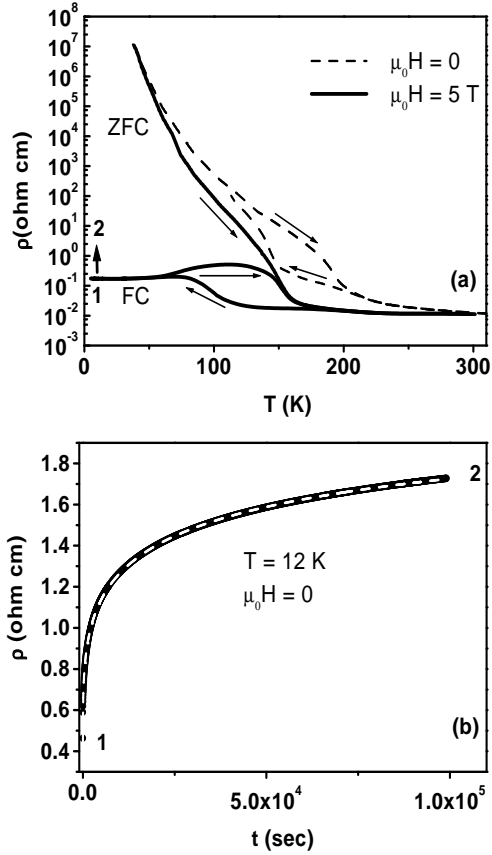,width=8.5cm,height=12.5cm,clip=}
}
\caption{(a) The temperature dependence of the resistivity of La$_{0.5}$Ca$%
_{0.5}$MnO$_{3}$ in magnetic field $H=0$ and 5 T. The thermal and magnetic
history of the sample is denoted as ZFC for zero field cooling and FC for
field cooling. The direction of the temperature change is shown by the
arrows. (b) The time dependence of the resistivity of La$_{0.5}$Ca$_{0.5}$MnO%
$_{3}$ at $T=12$ K taken after the sample was cooled in a magnetic field of
5 T then decreased to zero. The white dashed line is the fit described in
the text.}
\label{fig1}
\end{figure}


When La$_{0.5}$Ca$_{0.5}$MnO$_{3}$ is cooled in zero applied magnetic field
and then 8.5 T is applied, the resistivity decreases by many orders of
magnitude, but does not reach the value obtained when cooling in 8.5 T (see
Fig. 2a). Following this procedure (Path 2), the resistivity of the sample
continues to decrease as shown in Fig. 2b. This time dependence can also be
fit with a stretched exponential dependence (Eq. (1)) with the following
values of the parameters: $\rho _{1}=10.6\pm 0.2$ ohm-cm, $\rho _{2}=109\pm
2 $ ohm-cm, $\tau =206\pm 20$ sec, and $\beta =0.192\pm 0.004$. The time
dependence of the magnetization corresponding to this process (relaxation
\begin{figure}[tbp]
\centerline{
\psfig{figure=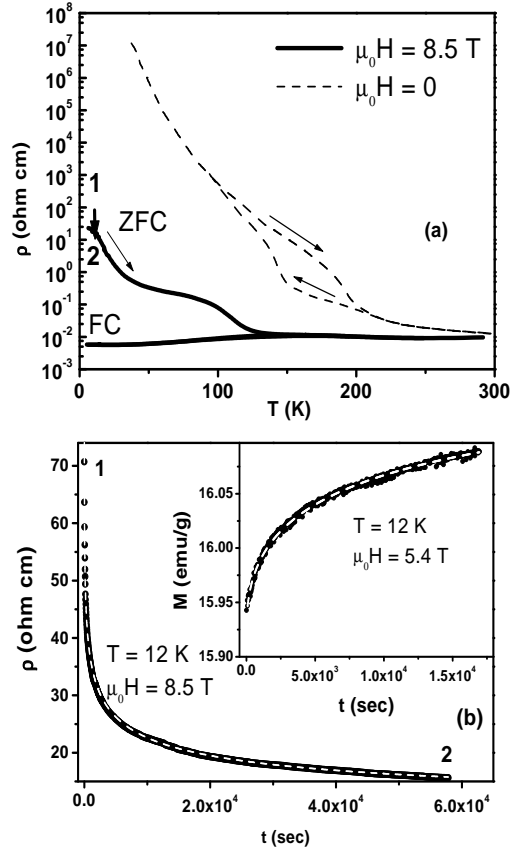,width=8.5cm,height=12.5cm,clip=}
}
\caption{(a) The temperature dependence of the resistivity of La$_{0.5}$Ca$%
_{0.5}$MnO$_{3}$ in magnetic fields $H=0$ and 8.5 T. (b) The time dependence
of the resistivity of La$_{0.5}$Ca$_{0.5}$MnO$_{3}$ at $T=12$ K taken after
ZFC and application of a 8.5 T field. The inset shows the time dependence of
the magnetization taken after ZFC to 12 K and application of 5.4 T field.
The white dashed lines are the fits described in the text.}

\label{fig2}
\end{figure}
after Path 2) is shown in the inset of Fig. 2b. These relaxation data were
taken after the magnetic field applied to the ZFC sample had reached 5.4 T.
The time dependence of the magnetization also follows a stretched
exponential dependence:

\begin{equation}
M(t)=M_{2}+(M_{1}-M_{2})exp[-(t/\tau )]^{\beta }
\end{equation}
where $M_{1}=16.176\pm 0.007$ emu/g, $M_{2}=15.921\pm 0.002$ emu/g, $\tau
=12500\pm 1000$ sec, and $\beta =0.45\pm 0.01$.

The relaxation curves shown in Figs. 1b and 2b are representative curves -
similar behavior occurs for different magnetic fields and temperatures below
the CO temperature. The values of $\tau $ range from 200 sec to 7250 sec for
the 8.5 T measurements and from 4350 sec to 34700 sec for the 5.4 T
measurements at different temperatures. The values of $\beta $ are found to
be between 0.192 and 0.45 for relaxations under different conditions. The
fact that the magnetic field cannot be removed (or switched on) instantly
(we used the rate 1.3 T/min) may influence the parameters $\tau $ and $\beta 
$ differently at different temperatures and magnetic fields. These
variations in parameter values are not understood at present and a more
detailed investigation will be required.

The time dependence of the resistivity taken after the Path 3 procedure also
follows a stretched exponential form. Moreover, another CO compound, La$%
_{0.47}$Ca$_{0.53}$MnO$_3$, exhibits similar behavior. We have not observed
sudden jumps in the resistance after any of our thermal and magnetic
treatments (Path 1, Path 2, and Path 3) in contrast to the jumps in the
resistance observed in another CO system, Pr$_{0.67}$Ca$_{0.33}$MnO$_3$ \cite
{Anane}.

In order to explain these memory and relaxation effects, we consider the
free energy of our system to be of double-well form as a function of the
magnetization (Fig. 3a), as had been proposed in Ref. 3 for the
interpretation of the temperature dependence of the critical field and the
hysteresis in Nd$_{0.5}$Sr$_{0.5}$MnO$_{3}$. The ZFC La$_{0.5}$Ca$_{0.5}$MnO$%
_{3}$ is in the AFM insulating (AFM-I) state because the lowest minimum of
the free energy in zero magnetic field corresponds to the AFM-I state.
However, in a magnetic field the FM metallic (FM-M) state becomes the most
energetically favorable state and the AFM insulating state no longer
corresponds to the lowest minimum in the free energy (Fig. 3a) (in magnetic
fields high enough to destroy the CO completely, the FM-M state becomes
perhaps the only free energy minimum). According to this picture the FC La$%
_{0.5}$Ca$_{0.5}$MnO$_{3}$, which is in the low resistivity state,
corresponds to the lowest minimum (Fig. 3a, $H>0$, FM-M state). After the
magnetic field is reduced to zero, the FM-M state is no longer the lowest
minimum. Nevertheless, since the AFM-I and FM-M states are separated by an
energy barrier, the system remains in this local minimum (the FM-M
metastable state) and slowly relaxes with time to reach the lowest AFM-I
minimum. This model qualitatively explains why the La$_{0.5}$Ca$_{0.5}$MnO$%
_{3}$ remains in the lower resistivity state even if the magnetic field has
been removed. A similar explanation applies to the ZFC sample put in a
magnetic field, but one not high enough for complete melting of the charge
ordering.

If the free energy of the system has only one potential barrier $U$ which
separates two minima, the relaxation should have only one relaxation time $%
\tau$ and exhibit an exponential time dependence (Debye relaxation with a
time constant $\tau \propto exp(-U/k_{B}T)$). In contrast, if there is a
distribution of the barriers and consequently a distribution of the
relaxation times, the relaxation can have a stretched exponential form
(Kohlrausch law \cite{Kohlr}): $q(t)=q_{0}exp[-(t/\tau )^{\beta }]$, where $%
q(t)$ is a relaxing quantity and $0<\beta <1$. This type of relaxation has
been observed for many complex and strongly interacting materials \cite{sg}.
\begin{figure}[tbp]
\centerline{
\psfig{figure=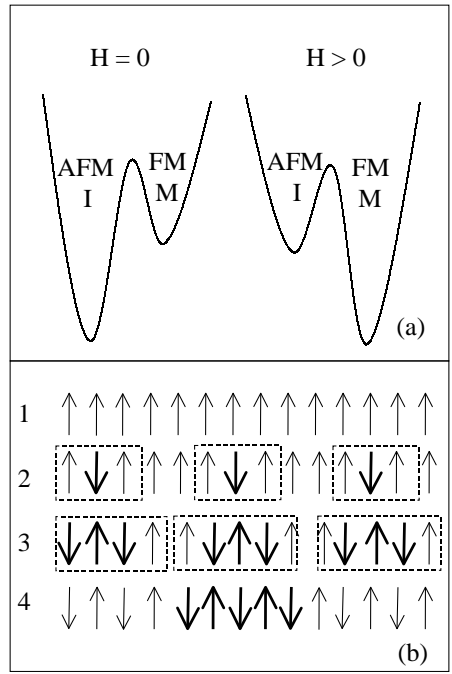,width=8.5cm,height=11cm,clip=}
}
\caption{(a) Schematic dependence of the free energy versus magnetization,
with and without magnetic field, with the two local minima corresponding to
the ferromagnetic metallic (FM-M) and antiferromagnetic insulating (AFM-I)
states. (b) Schematic picture of the changes occurring in the magnetic
system for the FC sample after the magnetic field is switched off, as
described in the text ($1\rightarrow 2 \rightarrow 3 \rightarrow 4)$. The
spins which change direction with respect to the previous step are shown in
bold. The dashed lines mark the regions within which AFM ordering is
established.}
\label{fig3}
\end{figure}

For glassy relaxation a model of hierarchically constrained dynamics has
been suggested \cite{Anderson}. According to this model, the relaxation of
the Kohlrausch type arises from many sequential correlated activation steps
with a hierarchy of relaxation times, from fast to slow. In our case it
could correspond to the appearance of small regions of the AFM phase in the
FM matrix (or small regions of the FM phase in the AFM matrix depending on
the thermal history) and the gradual growth of these regions to reach the
most energetically favorable state. A one dimensional illustration of this
process is shown schematically in Fig. 3b. If the sample is cooled down in
magnetic field, we start from the FM-M state (row 1 in Fig. 3b). When the
magnetic field is switched off and the system is no longer in the lowest
energy minimum (corresponding to AFM-I), it starts to change its state in
the following way. First, single spins change direction forming an AFM
ordered groups of three spins (Fig. 3b, row 2), which corresponds to the
smallest relaxation time (overcoming the smallest energy barrier). Next,
larger groups of spins change direction (Fig. 3b, row 3) resulting in larger
clusters of the AFM ordering, which corresponds to a longer relaxation time.
Finally, when even larger clusters move, the complete AFM ordering is
achieved (Fig. 3b, row 4). A similar process can occur in a three
dimensional system which has a large number of relaxation times
corresponding to the distribution of the energy barriers. Since the AFM
phase is insulating and FM phase is metallic, the resistivity reflects the
changes in the magnetic system. The change in the resistivity is more
appreciable than the change in the magnetization, perhaps, because the
percolative paths for the current could be opened (or closed) during the
process described above resulting in larger changes in the resistivity than
in the magnetization and different $\tau $ and $\beta $ for the relaxation
of magnetization and resistivity.

To summarize, we have measured the time dependence of the resistivity and
magnetization of La$_{0.5}$Ca$_{0.5}$MnO$_3$ and La$_{0.47}$Ca$_{0.53}$MnO$%
_3 $ ceramic samples after different thermal and magnetic treatments at
various temperatures and magnetic fields. A stretched exponential relaxation
of the resistivity and magnetization has been observed in the metastable
state. The relaxation process is due to the gradual change from the AFM-I
phase to the FM-M (or FM-M to AFM-I). We explain this relaxation behavior by
a model with a hierarchical distribution of relaxation times from fast to
slow.

The authors thank C. Lobb, R. Ramesh, P. Fournier, S. Bhagat and A. Biswas
for helpful discussion. This work was supported in part by a NSF MRSEC Grant
at the University of Maryland (No. DMR96-32521).

\end{document}